\documentclass[review]{elsarticle}

\usepackage{hyperref}
\usepackage{xcolor}

\journal{Journal of \LaTeX\ Templates}

\begin{document}

\begin{frontmatter}

\title{From impact refugees to deterritorialized states: foresighting extreme legal-policy cases in asteroid impact scenarios}

\author[A1]{Elisa Simó-Soler\corref{mycorrespondingauthor}}
\cortext[mycorrespondingauthor]{Corresponding author}
\ead{elisa.simo@uv.es}

\author[A2,A3]{Eloy Peña-Asensio}

%\author{Elisa Simó-Soler, Eloy Peña-Asensio}
% \fnref{myfootnote}



\address[A1]{Universitat de València (UV), Valencia, Spain}
\address[A2]{Universitat Autònoma de Barcelona (UAB), Bellaterra, Catalonia, Spain}
\address[A3]{Institute of Space Sciences (CSIC), Cerdanyola del Vallès, Catalonia, Spain}

\begin{abstract}
Throughout recorded history, humans have crossed national borders to seek safety in nearby countries. The reasons for displacement have been generated by phenomena of terrestrial origin, but exposure to unexpected extra-terrestrial threats poses a different scenario. An asteroid impact warning implies a change of paradigm which would represent a historic precedent. In this regard, the analogies with natural disasters must be considered, along with multiple possible scenarios, and legal aspects related to: a) the legal framework to regulate this situation; b) the action and responsibility of the states; and c) the definition of \textit{impact refugee} and the reconfiguration of traditional concepts such as deterritorialized states. In addition, the decision-making process and the actors involved must be led by a cooperative effort to improve international law. These new circumstances should be established with a consideration of inequalities between the states, and an aim of protecting humanity through democratic solutions using the safest, most effective techniques. 

\end{abstract}


\end{frontmatter}

\section{Introduction}

Migratory movements, as currently understood, have been caused by armed conflicts, political persecution, poverty, hunger, climate change effects, and/or natural disasters, all of which have a terrestrial origin. However, with a cosmic impact threat, the source of danger is outside the atmosphere, there being no prior human activity that motivated the displacement. A real impact hazard scenario would imply a change of paradigm, which would represent a historic precedent \cite{schmidt2018planetary}.

Three well-known events provide a deeper understanding about the destructive power of such a collision: the Chicxulub impact that triggered a mass extinction 65.5 million years ago \cite{schulte2010chicxulub}; the explosion over Tunguska in 1908 which could have resulted in massive fatalities, had it occurred over a large metro area \cite{chyba19931908}; and the most recent and unexpected event that took place in Chelyabinsk on 15 February 2013, causing material and human damage \cite{borovivcka2013trajectory, popova2013chelyabinsk}. It is worth noting that the latter occurred at a time when the sky was already being monitored, and state-of-the-art near-Earth object (NEO) survey technologies were available.

According to NASA’s Near-Earth object survey and deflection analysis of alternatives \cite{national2007near}, a new Tunguska-like event could cause approximately half a thousand fatalities, and an object measuring 140 meters in diameter striking the Earth would increase the number of victims to 50000, with the probability of this happening in the next 50 years being $\sim1\,\%$. An impact of an asteroid as small as 600 meters long could cause 5 million human casualties. The estimated frequencies of thousands or millions of years for such events may seem low on a human scale. However, this is only a statistical average of random phenomena, and the world may face an impactor crisis of regional proportions at any time. Moreover, it is estimated that two thirds of NEOs large enough to cause regional damage have not yet been detected \cite{mainzer2020future}.
 
In the last few years, asteroid impact mitigation proposals have been proliferating (e.g., Kinetic Impactor, Ion Beam, Gravity Tractor, Nuclear Device…), suggesting promising approaches to avoid potential catastrophic asteroid–Earth collisions \cite{weisbin2015comparative}. Some of these options, explosives in general (because of the risk of fragmentation) \cite{BruckSyal2013} and nuclear in particular (because of the risks of nuclear war and/or violent conflict) \cite{baum2019risk}, may be counterproductive. A correction of asteroids’ orbits is a candidate strategy to prevent impacts with the Earth, especially the Kinetic Impactor as evidenced by the interest in various experiments being carried out such as \textit{DART} \cite{cheng2015asteroid} and \textit{HERA} \cite{sears2004hera}. However, many unknowns in the structure, composition, and mechanical response of the asteroid target can make it difficult to predict the exact deflection outcome \cite{syal2016deflection, tanbakouei2019mechanical}. In addition, responsiveness will be constrained by the time of discovery \cite{greenstreet2020required}, which is a random factor. This lack of knowledge, together with the inherent inaccuracies of this type of astrodynamics calculations \cite{Rumpf2020}, turn the collision threat context into an unpredictable scenario.

\section{Collision threat challenges}

This uncertainty contains a direct reflection on the decision-making process based on a two-fold approach: policy and regulation. On one hand, the complexity of predicting the impact point on the risk corridor could hinder the discussion of whether to deflect or not. In addition, due to lack of time or technological capability, the physical properties of an asteroid on a collision course with the Earth could present large uncertainties and, therefore, deflection outcomes as well. On the other hand, the state of science could change the political scenario, being directly related to technological development at the time of potentially hazardous asteroid (PHA) discovery. Significant improvements in deflection capabilities would entail a radical modification of the legal framework: the enhancement of laser or ion beam techniques \cite{brophy2018characteristics}, the shortening of the response time by dramatically improving the spacecraft launches and the possible mission payloads \cite{barbee2018options}, or the development of exotic methods, such as antimatter weapons or solar sails \cite{chapman2015facing}. This analysis leads us to assess circumstances such as these outlined below. 
 
Technological dependence can lead to difficult dilemmas. The state could only partially divert the asteroid changing the target region, having an opportunity to prioritize either an industrialized or high density-populated area. With greater control of the deflection, the strategy could then become discretionary, forcing the establishment of conditions to prevent conflicts. For example, no strategic geopolitical interests should be involved in the choice. Ironically, if we do not develop deflection technology we cannot defend against an incoming asteroid, but, if we do develop this capability, someone could deliberately misuse it with potentially apocalyptic results \cite{sagan1994}. Similarly, a more pertinent issue is the so-called Real Deflection Dilemma, which states that a planetary defense mission necessarily places otherwise non-threatened lives, and tangible and intangible assets, in jeopardy \cite{schweickart2004real}. The preference for a collision in a territory uninhabited by humans could be considered, but what about other living beings? What about the implications for climate change if the asteroid impacts the Amazon or Antarctica? What about if the asteroid impacts the Heritage of Humanity or the cultural-religious heritage of a community? 

Along with uncertainty, there is another element that characterizes this impact threat context: collaboration. We have already learned that international cooperation is necessary to deal with cross-border problems \cite{taylor1987possibility, wilson2012companion, hoffmann2011climate, hiro2014war, luengo2020artificial}. In the face of globalized systems of insecurity where there is no singular winner, but humanity as a whole would benefit, the postulates of security cosmopolitanism become relevant.

This approach reconceptualizes transnational security through the ethos of human solidarity, dignity, and agency, and promotes a theoretical frame “in which the security of all states and all human beings is of equal weight, in which causal chains and processes spread widely across space and through time, and in which all security actors bear a responsibility to consider the global impact of their decisions” \cite{burke2013security}. An asteroid deflection would need to be implemented from international coordination and global networking (scientific community, technological resources, techniques, budget, and decision-making processes) \cite{packer2013policy}. 

Given that only several states and non-governmental bodies would have the capacity to respond, we refer to a partnership based on solidarity. According to Macdonald, “solidarity is first and foremost a principle of cooperation which identifies as the goal of joint and separate state action an outcome that benefits all states or at least does not gravely interfere with the interests of other states. Solidarity, as a principle of international law, creates a context for meaningful cooperation that goes beyond the concept of a global welfare state; on the legal plane it reflects and reinforces the broader idea of a world community of interdependent states” \cite{macdonald1996solidarity}.

It would be desirable to place the expertise and knowledge of a few for the service of all humanity. However, this statement can only be read as a declaration of intent, since, as we shall see, there is no legal obligation to undertake it. Although it is not clearly defined in international law, the principle of solidarity has served as a basis for the recognition of certain rights in the paradigm of maintenance of international peace and security: the right to receive assistance, the right to offer it, and the right of access to the victims (e.g., natural disasters) \cite{wolfrum2010solidarity}.

While it is true that planetary defense would require normative development, it is possible to articulate its use at least from its definition as a fact or condition \cite{carozza2014solidarity}, even in relation to third generation rights. The PHA reality urges us to compel necessity in the development of international solidarity \cite{wolfrum2010solidarity}. {Before grappling with these issues and the related legal-policy problematics, we employ the climate change framework as a gateway to analyze the collisional threat context and its possible consequences.}

\section{Overview of historical analogues}

According to IFRC, “a disaster is a sudden, calamitous event that seriously disrupts the functioning of a community or society and causes human, material, and economic or environmental losses that exceed the community’s or society’s ability to cope using its own resources. Though often caused by nature, disasters can have human origins” \cite{IFRCweb}. The Secretary-General Report 2019 states: “International cooperation on humanitarian assistance in the field of natural disasters, from relief to development”, “disaster displacement has devastating effects on people and their livelihoods. Between 2008 and 2018, disasters triggered more than three times the internal displacement than conflict and violence. […] Weather-related hazards account for more than 87 per cent of all disaster displacement, and their risk and impact are expected to be aggravated by climate change. A global average of more than 17 million people is at risk of being displaced annually by floods alone, with more than 80 per cent of them living in urban and peri-urban areas” \cite{assembly2019international}.

Applying this definition, and given the current prevalence of natural disasters, it is possible to establish a comparative framework to shed light on the search for solutions to a possible colliding asteroid with devastating consequences. Wars, economic blockades, colonization, the plundering of raw materials, famine, and poverty are examples of direct human causes for displacements. Although they may leave a country uninhabited, and provide asylum and refugee status in certain cases (in others, apathy and indifference prevail), in the realm of environmental catastrophes, it is more useful to seek historic analogues taken as reference. 

A review about natural disasters can help policy makers to design the evacuation strategy and to anticipate complex issues. First of all, natural disasters and interplanetary collision hazards share a wide margin of uncertainty.  Hurricanes, typhoons, tsunamis, tornadoes, earthquakes, floods, volcanic eruptions, wildfires, landslides, avalanches, and heat waves can occur suddenly at any time, in any place in the world. This quality of unpredictability is also present in asteroid impacts. However, not all are sudden-onset emergencies. Like drought, desertification, and glaciation, an asteroid collision can have slow onset, and can be forecast even decades beforehand. Another common point is that cause and blame cannot be easily and directly attributed to human actions, or the effect of the policies applied \cite{spiegel2005differences}.

Secondly, in the aftermath of a natural disaster, particularly in the event of an asteroid impact, the affected population must be provided with basic needs such as water, sanitation, food, shelter, health care, and protection. Each requires additional funding, immediate and long-term effective aid, and data to make decisions and evaluate interventions \cite{spiegel2005differences}.
Thirdly, during climate disaster management, there are two elements that need to be kept in mind when transferring the context to the impact of an asteroid. On one side, the vulnerability of women and children make them potential victims of sexual exploitation and gender-based violence. In addition, mental health problems due to the emotional shock must be addressed. On the other side, NGOs are the first barrier of protection on the ground. When drafting the planetary defense action plan, the third sector must be taken into consideration as a key active agent from the beginning and not only to supply institutional weaknesses and assistance gaps \cite{spiegel2005differences}.

Climate change nature could be described as a hyperobject, massively distributed in time and space, multidimensional, and with apparently chaotic behavior \cite{morton2013hyperobjects}. These characteristics make climate change incomparable to an asteroid collision context. Despite this, the novel challenges presented by the climate change crisis can be useful to face the problems in the context of asteroid being on a collision course with the Earth. Complex but fundamental issues such as the redefinition of citizenship-sovereignty-territory, diffuse liability for damage, or migrant policy to prevent stateless citizens, have been considered \cite{marshall2020climate, yamamoto2017migration}. Currently, the international community must deal with the matter of relocation due to extreme adverse weather conditions. In particular, Canada’s coastal Inuit communities are suffering from melting permafrost and small islands like Kiribati, Tuvalu, Maldives and Marshall Islands could become submerged in the near future. To avoid inhabitants being placed in a legal limbo or even becoming stateless, a binding instrument with bilateral or regional agreements should be signed, and soft law norms should be adopted, to guarantee the rights of displaced people \cite{ yamamoto2017migration,atapattu2014climate}.

Measures implemented to try to reduce the harm brought on by natural disasters can be used as models. For instance, despite climate change being a multifactorial phenomenon, the human contribution is evident, especially in the context of more developed countries \cite{cook2016consensus}. May financial compensation be demanded from them? Could a legal claim against private polluters companies be possible? Whose responsibility is it to lead and develop the relocation plan? \cite{atapattu2014climate}

To circumvent a “wait-and-see” policy, as is mostly happening with climate change, we propose as key points foresight and preparedness. The debate initiated on states without territory and climate refugees could be transposed, at least partially, to the asteroid impact scenario.

\section{Hypothetical impact scenarios}

\begin{figure}[htb]
\centering
\includegraphics[width=\textwidth]{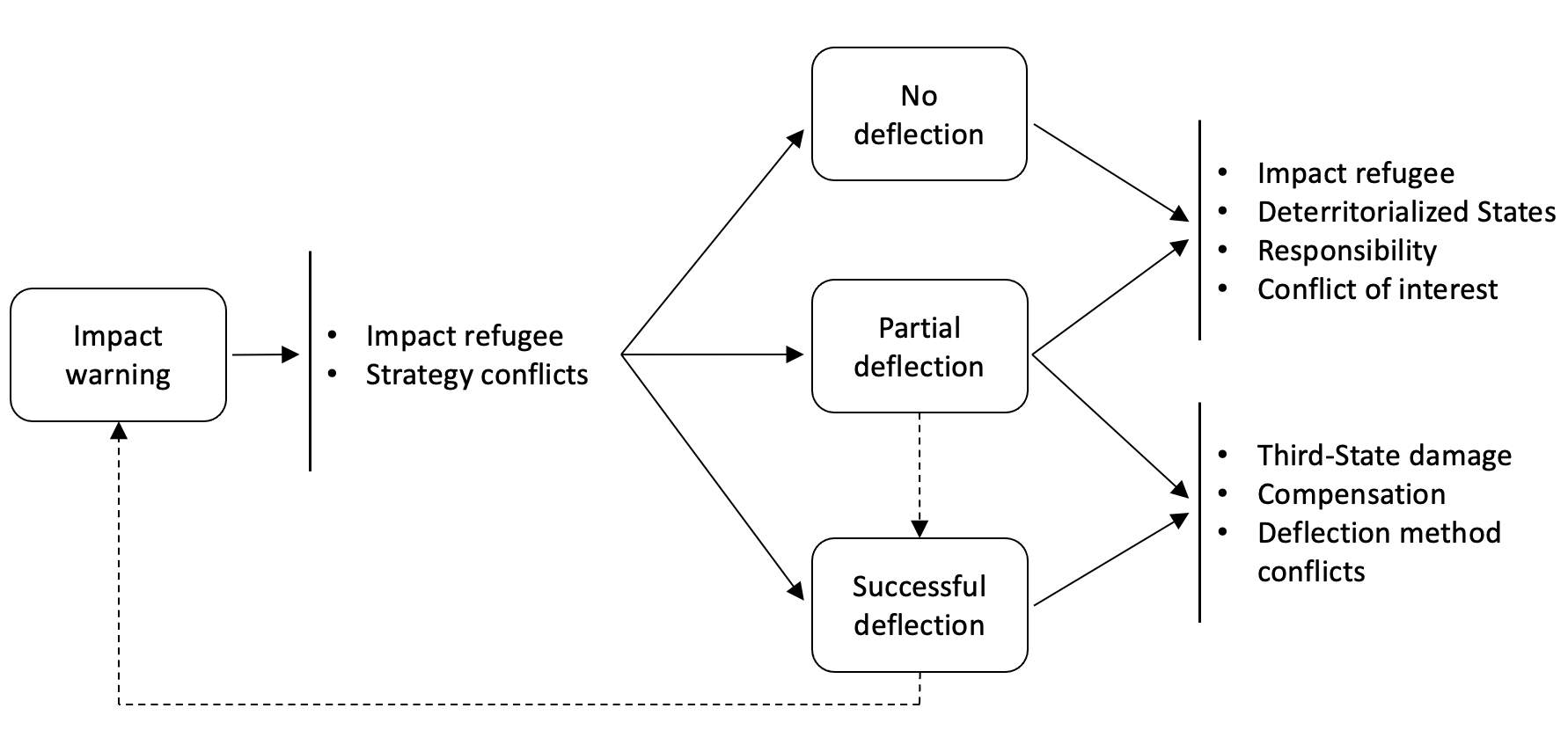}
\caption{Schematic diagram of asteroid impact threat scenarios and possible consequences.}
\label{fig:ScenariosDiagram}
\end{figure}

In this section, we examine some political fiction situations with the purpose of exploring as many options as possible to lay out legal groundwork options beforehand, since it is rational to follow a “prevention is better than cure” strategy that could mark governmental action for the protection of humanity ensuring the least damage as possible. We have set up four frameworks, with both individual and shared elements, which define multiple scenarios. The four hypothetical cases are impact warning, no deflection, partial deflection, and successful deflection (see Figure \ref{fig:ScenariosDiagram}).

\subsection{Asteroid impact warning}

As stated above, there is a high degree of uncertainty in determining the probability of an impact. Despite (or perhaps because of) this, it may be necessary to take action (and with prior legal agreements) as if the impact were certain, in order to prepare for the worst-case scenario with sufficient time.

The impact threat could trigger a migration phenomenon to low-risk regions, or even a global migration, to the opposite impact side of the Earth. It could be spontaneous or institutionally coordinated and temporary or permanent, but in any case, it should be supported by a legal framework: the impact refugee (discussed later in section \ref{ImpactRefuDeterStat}). 

Different conflicts of strategy would arise for many reasons: lack of technological capacity, general cost-loss balance sheet, or faith in destiny. Specific scenarios are described below:

- A state employs a dangerous and questionable technique that could generate greater harm such as the use of Nuclear Explosive Device \cite{schweickart2004real}. The first risk would simply be a launch failure resulting in radioactive pollution on the ground or at low altitude. A modest yield standoff burst or an unexpected response might not destroy the asteroid but produce fragments, which could be gravitationally bound and still on a collision path \cite{syal2013limits}. This situation would hinder further deflection attempts and threaten new regions instead of eliminating the danger \cite{Rumpf2020}. Moreover, the most optimal method of deflection (e.g., high yield standoff bursts) could exacerbate nuclear tensions \cite{schmidt2019political}. For this reason, it should be relegated as \textit{ultima ratio} \cite{baum2019risk}. Following this example, the threatened state could decide to discard the nuclear deflection. The debate then arises about the extent of national sovereignty and the non-existent obligation of states to assist others \cite{drubehaddaji2020legal}. 

- To maximize the operational control and ensure the highest responsiveness, states could nationalize key companies not only for planetary defense, but also for economic recovery. Without this private-public conversion, a satisfactory method outcome (thanks to private corporations) could force compensations or special treatment. 

\subsection{No deflection}

The inactivity of the threatened state or a failed mitigation mission may result in a catastrophic event. The territory could be devastated and its infrastructure (housing, parliament, hospitals, courts, police stations, businesses, and offices) destroyed. Depending on the estimated potential damage, the consequent migratory phenomenon could be temporary or inevitably permanent \cite{ferris2010natural, atapattu2014climate}.

The existence of an uninhabitable territory, or its direct physical extinction, suggests the reconfiguration of the nation-state model under the paradigm of deterritorialized states (addressed below in section \ref{ImpactRefuDeterStat}).

In the third case, if actors (states, companies or international organizations) decide to deflect the asteroid with a failed outcome, there would be no current legal obligation to assist the affected state. In principle, ignoring the request for assistance would not violate any international norm. Consequently, if we wanted to forestall or minimize the hardships that would likely result from ignoring the distressed populations, it would be necessary to determine beforehand an accountability regime to delimit the legal margins of inaction.

An engagement of a third entity would be considered based on a cost-benefit analysis with conflict of interest. A state may deny assistance because of global market competition, or, it may help to pursue its geopolitical purposes. In this context, a threatened state without sufficient technological capacity could face abusive negotiations for assistance. Indeed, an asteroid impact scenario places the question about the equal value of all human and non-human lives, as well as the protection of the planet biosphere, at the center of the debate \cite{schmidt2018planetary}, especially with the rise of anti-speciesist, ecologism, and environmentalist perspectives \cite{horta2010speciesism, pendergrast2016environmental}.

\subsection{Partial deflection}

Two sub-scenarios are envisaged within this framework. First, the asteroid is partially deviated due to limited mitigation capacity or failed mission outcome, causing damage in a region not originally at risk. All the elements are shared with the previous framework: the migration phenomena, the states without territory, the call for cooperation, the responsibility for a strategy failure, and the role of conflicting interests. 

Second, a succession of unilateral deflections between nations to minimize their risks could cause the object to ultimately impact a third state, different from the one initially threatened. For example, this would be the case where a first state deflects the asteroid, then a second state deflects it, and the asteroid finally collides with an ocean, causing a tsunami to strike a third state's coast that was not previously threatened. The consequences should be analyzed: the penalization of states and the claim for compensation from the damaged state. Geopolitical interests and lack of technological capacity, as well as disagreements in the applied methods, are again at the forefront.

\subsection{Successful deflection}

A successful asteroid deflection would be the optimal outcome. However, this would not avoid possible further close encounters typical of PHA orbits \cite{giorgini2008predicting}. Therefore, impact hazard may arise later, threatening a new region of the Earth. This frame contains the same components as outlined above, but with some particularities. A third state that was not initially threatened might now be endangered. Under a later threat of impact, two legal consequences reemerge: the claim for compensation by the damaged state, and the reward to the state or entity that has successfully conducted the defensive strategy. The case of a post-deflection dangerous re-encounter involves an additional challenge, since the degree of predictability needed to prevent it (or plan for it) is far beyond our current knowledge.

Regarding the defense methods, the asteroid may be successfully deflected, but in violation of international law. This case should be justified by the triple canon of distress, necessity, and consent \cite{drubehaddaji2020legal}.

Generating a body of law requires an in-depth knowledge of the specific factual assumption. Extrapolating current knowledge about the techniques and possible outcomes is therefore a first step for foresight and prevention. The more realistic scenarios included in the regulation, the higher the degree of protection.

\section{From impact refugees to deterritorialized states}
\label{ImpactRefuDeterStat}

From the hypothetical scenarios, and using the controversial \textit{climate refugee} as a benchmark, we anticipate the figure of \textit{ impact refugee} as a key legal-political concept in a collision threat context. 

The U.N. High Commissioner for Refugees (UNHCR) and the International Organisation for Migration (IOM) do not recognize, and refuse to use the term, \textit{ climate change refugees}. One of the main reasons for its rejection is that the 1951 Refugee Convention contains a \textit{numerus clausus} of circumstances that determine refugee status and that do not allow for an expansive interpretation  \cite{williams2008turning, biermann2008protecting}. The alternatives are to add a protocol on climate-induced migration to the United Nations Framework Convention on Climate Change, or, to create a new legal and political regime specifically for the needs of \textit{ impact refugees}: recognition, protection and, where necessary, compensation and resettlement\cite{hartmann2010rethinking, veening2014climate}.

Similarly, when proposing an environmental refugee status, several aspects need to be addressed to its configuration. First is the reason for migration (linked to the possible scenarios outlined above). Second is the period of migration (temporary or permanent). Third is the migration's dimension (with or without transboundary movement) \cite{biermann2008protecting}. Each of the possibilities will involve specific action on the part of the threatened state, the technologically developed states, and the international community.

Linked to this, we could consider the existence of states without territory, or deterritorialized states. Underlying this potential reality is the concept of \textit{ex-situ} nationhood proposed by Burkett: “[It] would be a status that allows for the continued existence of a sovereign state, afforded all the rights and benefits of sovereignty amongst the family of nation-states, in perpetuity. It would protect the peoples forced from their original place of being by serving as a political entity that remains constant even as its citizens establish residence in other states. It is a means of conserving the existing state and holding the resources and well-being of its citizens -in new and disparate locations- in the care of an entity acting in the best interest of its people” \cite{burkett2011nation}. The remote possibility of a territory disappearing completely or becoming uninhabitable, with consequent large-scale migration, involves an effort of conceptual flexibility to re-think the modern nation-state system, and to redefine essential elements such as citizenship and state sovereign authority \cite{marshall2020climate}.

Accordingly, solutions to this situation should be sought: could the international community recognize states without territory? Could the formula of government in exile be adapted? Would the population be relocated to another territory or dispersed in several states? Could a state cede part of its territory to another one? Would it remain the sovereign government of a devastated territory? Would state-reconstruction be financially supported by the international community? \cite{douglas2017framework} 

These types of question have been discussed in the Small Island Developing States framework \cite{woodward2019promoting}. To retain their sovereignty, nations ex-situ based on a political trusteeship system; governments-in-exile or deterritorialized states have been recommended by legal scholars as possibilities in the international law realm. The sooner appropriate measures and governance mechanisms are defined, the better position states and populations will be in.

Some caveats related to the actors involved in planetary defense are shared with climate migration. The five permanent members of the UN Security Council (the United States, China, Russia, France, and the United Kingdom, who are some of the largest emitters of greenhouse gases) overlap roles, and at the same time, would be the leaders in climate catastrophe response \cite{hartmann2010rethinking}. 

In deflection missions, the same paradox occurs. These same states have the technological capacity, and simultaneously, have the right to veto. This translates into an excessive concentration of power, which requires finding tools to democratize the oligopoly of planetary defense. This would be complicated further when the inherent right of individual or collective self-defense could be impaired if a state does not have the technological capability to prepare the deflection. In case of a veto, the threatening state could trigger Article 51 from the UN Charter, or explain the action as necessary circumventing the veto and still be legal. There are two hypothetical situations that are worth consideration.

Firstly, the right to veto could be understood as the right of individual self-defense if the deflection strategy potentially affects one of the five permanent members of the UNSC. Secondly, discrepancies in hazard assessment could lead to divergent positions whether a planetary defense action is justified or not, considering some members when the impact consequences are assumable (lower in cost than deflection, but what about the subjectivity of the socio-cultural value?) or that probability impact estimation does not require mitigation missions (where does the threshold lie?) \cite{drubehaddaji2020legal}. Moreover, in these approaches, there are also spurious geopolitical interests that come into play. As Schmidt argues, “we can say that all people tend to want to save the world, but the question is how big is the one – their one – they are saving? Saving humanity can for some be leverage for domestic political objectives”. A statement that is followed by the tough recognition that “there is no reason why we should assume that the \textit{modus operandi} would change in planetary defense” \cite{schmidt2018planetary}. 

In this respect, tackling the procedure and decision-makers (assuming that the common good should underpin the activity of states) becomes a priority. The planetary defense community should consider the emerging problem of the power politics game due to the absence of a robust legal framework, and the disturbance this may pose to the exercise of democracy. Without this assurance, “singular defense activities may become a destabilizing factor, trigger arms races, and destroy the balance between leading powers, not only in space but also on Earth” \cite{schmidt2018planetary}.
 
Although certain issues have already been addressed \cite{drubehaddaji2020legal}, it is important to point out some observations. Even though these are strictly political aspects, they should not be separated from the technological sphere and the progress made in science. It must be determined which institution would bear the decision. Will be the states with technological capabilities, the UN Security Council, or a specific institution created \textit{ad hoc}? An \textit{in extremis} problem-solving action could be unilateral? If the purpose is to democratize the decision-making process, would the referendum be a valid instrument? In that case: which population would be participating; what would the consultation be about (on the decision to act or not, on the method, on assistance to a third state?); how would the question be formulated, and would there be time to do it? These are, indeed, thought-provoking queries that need to be agreed upon in advance to avoid conflicts that will be aggravated by the immediacy of the decision and the damage of failure or inaction.

The preliminary study leads to the conviction that these new circumstances must be faced keeping in mind the inequalities between states, and protecting the most vulnerable groups. Policymakers and legislators must assume that an Earth-bound asteroid places the entire planet at risk. Anticipatory actions should be taken from the standpoint of international law and with the safest techniques. Particular emphasis should be given to the development of democratic solutions to protect humanity from an asteroid impact.

\section{Legal framework and proposals}

As a general rule, legal dynamics follow a pattern: Society and its demands precede law; thus, rules respond to needs. This can be seen in many and varied fields, from lowering taxes and recognition of gay marriage, to access to education and the banning of single-use plastics. On the issue of planetary defense, however, the process is the other way around. Science and law must anticipate the emergence of the factual assumption. It would be unimaginable to wait for a collision before regulating this scenario.

In addition, when several legal interests need to be protected, correct balance is required. Examining, on one side, the means to guarantee one right and, on the other side, the interference with others until a balance is reached. This results in an equilibrium point that fits within the sphere of values of each society and its rule of law. 

As a result, some legal aspects should be tackled considering geopolitical interests and the likely multipolar distribution of the world system. In a situation where technological capacity is held by major global powers, where private actors have more power than some states, and where the decision-making process is not well defined, the equilibrium of power may be altered.

In any case, it is a question of: a) studying the legal framework to regulate this situation \cite{ireland1967treaty}; b) defining the action and responsibility of the states \cite{rusek2008outer, bucknam2008asteroid}; and c) creating the new concept of \textit{ impact refugee} from the consolidated figure of refugee \cite{fischer2009border, loescher2008protracted, assembly1951convention, assembly1967protocol}. 

Currently, there are five main instruments ratified by the majority of states (the Outer Space Treaty, the Rescue Agreement, the Liability Convention, the Registration Convention, and the Moon Agreement). They cover issues concerning the obligation to inform and to act, the legality of planetary defense methods, the responsibility and liability regime, and related aspects such as space debris and the role of non-governmental entities. Soft law instruments are also useful to interpret and enforce international law\cite{soucek2016legal}. In addition to supranational standards, regional and national norms must also be kept in mind \cite{drubehaddaji2020legal}.

The main drawback is the limited time to make decisions and take action. It would be thus desirable to reach binding agreements broad enough to generate consensus, and to be adapted to the concrete collision circumstances. The more detailed the space activity, the greater the legal certainty for states and citizens. This is particularly relevant when actions may entail state responsibility, and there is a real danger to all forms of life.

In terms of the accountability system, the concepts of responsibility and liability must be distinguished. The first one refers to the “situation resulting from a violation of international law”, and the second one includes “the duty to compensate for damage”. The interplay between responsibility and liability can be summarized as follows: “a state may be held internationally responsible for a wrongful act although there is no material damage; a state may be held internationally liable for damage caused although it did not act wrongfully” \cite{drubehaddaji2020legal}.

Both parameters should be analyzed in detail, as well as the disclaimer. The legal and legitimate causes for which international law could be violated should be adapted to PHA context. As the SMPAG considers, “in practice it might prove difficult to conduct a comprehensive evaluation of whether in a concrete case the specific conditions for invoking circumstances precluding wrongfulness are met, such as whether the NEO threat constitutes a ‘grave and imminent peril’, whether a particular mitigation method is ‘the only way’ available to safeguard the interests endangered by the possible NEO impact and whether essential interests of other States are seriously impaired” \cite{drubehaddaji2020legal}. Likewise, the exemption from state responsibility for caused damage in the mitigation mission could be understood as an incentive for the nation, organism, or company that conducts the mission \cite{drubehaddaji2020legal}. The extreme (but plausible) case of imminent threat of mass extinction of all humankind could require rethinking about exceptions for a greater and unique good. This proposal could find its moral foundations in the theories of Cosmopolitan Responsibility. Those who are privileged (in this case because they have sufficient technological capacity or resources) could have the ethical duty to mitigate the asteroid impact by following the maxim that every member of the human community should be treated as equal. This would lead them to prevent structural injustice and seek relational equality \cite{heilinger2019cosmopolitan}. Once again, this is an exercise in adjusting the rules to an unprecedented factual situation.

Taking all of the above criteria into consideration, and assuming the differences between climate change and an Earth-asteroid collision, the institutional line to mitigate global warming could be given as an example of how to cope with a transnational problem that requires joint action.

The 2030 Agenda for Sustainable Development establishes that “the global nature of climate change calls for the widest possible international cooperation”. Problems from the globalized world call for cooperative solutions, as we recently saw in the response to the pandemic COVID-19. These could be extrapolated and adapted for an Earth-impacting asteroid scenario.

As the Secretary-General states in his report, “escalating disaster and climate risks require scaled-up investment in disaster risk management and building resilience. Effective early warning systems and evacuations are essential for saving lives and should form an integral part of preparedness and response strategies and sustainable development and adaptation efforts. Science, information and communications technology and big data can help to ensure that timely alert information reaches the last mile and triggers effective early action before disasters strike” \cite{ assembly2019international}. Despite the possible uncertainty, this group of measures could be adapted to the planetary defense context. 

Similarly, following in the footsteps of the fight against climate change, it might be desirable to hold an international working meeting to begin planning the planetary defense international guidelines. The Paris COP21 conference in 2015 highlighted states which were committed to cooperation, and in which an ambitious and universal climate agreement should be adopted. The United Nations Framework Convention on Climate Change is the primary international, intergovernmental forum for negotiating the global response to climate change. The Paris Agreement and the Kyoto Protocol are attempts to curb global warming. 

The International Asteroid Warning Network, the Planetary Defense Coordination Office, the Space Mission Planning Advisory Group, or the proposed Planetary Council \cite{schmidt2018planetary} could lead a similar conference where the possible creation of a future planetary defense treaty was debated. The migratory phenomenon should be tackled in this discussion to amend the Geneva Convention, or to create a specific regulation within or outside the planetary defense treaty.

Global collaboration is the cornerstone of the challenge we face as humanity. Paradoxically, a cosmic impact hazard could lead us towards the unification of humanity: a common enemy that does not distinguish between ethnicities, nationalities, or social groups; a shared threat that needs foresight and multilevel cooperation for maturing international law in order to provide an effective and desirable response.

\section{Conclusions}
 
This article is an overview of the political consequences of an impact asteroid threat from a techno-legal perspective. The technical approach establishes a realistic bodywork from the current state-of-the-art deflection, producing some scenarios for analysis. The legal approach allows for a complex and substantive study of legal and procedural issues and assessment of the feasibility of proposals. Geopolitical interests, human rights conventions, and democratic rules converge in the development of this political fiction exercise. The value of this work lies mainly not in its conclusions, but in its analytical framework for legal and political decision-making. 

In the absence of a precedent, we perform an extrapolation effort dismantling the traditional schemes, according to which social demands precede the law. Uncertainty and cooperation are characteristics for the design of the planetary defense strategy. Law should define a responsibility and liability system, and adapt to a new context. The impact threat scenario lays the groundwork for the creation of the \textit{ impact refugees} figure and the recognition of deterritorialized states. 

Although the obligation to protect concerns each state, not everyone would have the technological capability to deflect an Earth-approaching object. The concentration of technological and political powers cannot jeopardize the survival of a part of humanity.

The decision-making process must meet democratic standards and seek to protect the entire population without the distortion of geostrategic and ideological interests, governed by the principle of equality. The time to prepare for an asteroid impact scenario, first conceptually and then legally, is now.

\section*{Acknowledgement}

ESS acknowledges financial support from the Spanish Ministry of Science, Innovation and Universities (FPU 17/04436). The authors thank Professor Emeritus Dr. Joel Marks for the constructive comments that helped us to improve this paper.

\bibliography{mybibfile}

\end{document}